\documentclass{article}
\usepackage{amssymb,amsmath,amsfonts}

\begin{document}

\title{Mean-field theory of the Interaction of the Magnesium Ion with Biopolymers:\\ The Case of Lysozyme}
\author{Theo Odijk\\
Lorentz Insitute for Theoretical Physics\\
Leiden University\\
The Netherlands\\
E-mail: odijktcf@online.nl}

\maketitle

\begin{abstract}
A statistical theory is presented of the magnesium ion interacting with lysozyme under conditions where the latter is positively charged. Temporarily assuming magnesium is not noncovalently bound to the protein, I solve the nonlinear Poisson-Boltzmann equation accurately and uniformly in a perturbative fashion. The resulting expression for the effective charge, which is larger than nominal owing to overshooting, is subtle and cannot be asymptotically expanded at high ionic strengths that are practical. An adhesive potential taken from earlier work together with the assumption of possibly bound magnesium is then fitted to be in accord with measurements of the second virial coefficient by Tessier {\em et al.} The resulting numbers of bound magnesium ions as a function of MgBr$_2$ concentration are entirely reasonable compared with densitometry measurements.

\end{abstract}

\newpage

\section{Introduction}

We have a fairly good understanding of the way biopolymers interact with monovalent ions like Na$^+$ and Cl$^-$ (although there is now evidence that the electric fields of polyelectrolytes are so high that they influence the quantum mechanical properties of water \cite{1} which has obvious implications for charged biopolymers). On the other hand, the interaction with multivalent ions remains elusive. The magnesium ion at small concentrations, for instance, has a strong influence on the thermodynamic properties of DNA solutions as was established by Lerman {\em et al.} a long time ago \cite{2}. In the case of lysozyme, the Mg$^{2+}$ ion binds noncovalently to the positively charged protein but this happens at high concentrations of the cation \cite{3}. The binding appears to be corroborated in studies of the second virial coefficient $B_2$ of lysozyme in MgBr$_2$ solutions where a minimum was found at around 0.3 M \cite{4,5}.

The second virial coefficient of lysozyme in NaCl solutions was measured thoroughly by many experimental groups which allowed Prinsen and myself to establish the two parameters of the purported adhesive potential $U_{\rm A}$  between lysozyme spheres quite unambiguously \cite{6}. The potential is independent of protein charge and ionic strength so there is sound reason to hypothesize that it remains valid even when the salt is divalent like MgBr$_2$. As in Prinsen and Odijk \cite{6}, I solve the Poisson-Boltzmann equation perturbatively in order to compute the effective charge of the protein based solely on electrostatics. The magnesium ion is excluded from the lysozyme surface so that the effective charge is larger than nominal as has already been discussed by Tellez and Trizac \cite{7}. The object of this paper is to set up a self-consistent theory of the interaction of the magnesium ion with positively charged lysozyme. Because the binding constant of the ion is unknown a priori, I evaluate an actual effective charge as an adjustable parameter via the measurements of $B_2$ \cite{4,5} by letting the lysozyme spheres interact via the Poisson-Boltzmann equation and the adhesive potential $U_{\rm A}$. The resulting values of bound Mg$^{2+}$ as a function of Mg$^{2+}$ concentration are then compared with those established by densitometry \cite{3}. 

Tellez and Trizac already presented interesting numerical and analytical computations for spherical and cylindrical colloids in a 2--1 electrolyte (the cation is divalent whereas the counterion is monovalent) \cite{7}. Their analysis extends the previous multiscale method of Shkel {\em et al.} \cite{8} and is useful when $a \kappa \!>\! 1$ where $a$ is the radius of curvature and $\kappa^{-1}$ is the Debye screening length, as they showed numerically. Here, the objective is different: I solve the nonlinear Poisson-Boltzmann equation perturbatively for all $a \kappa$ where $a$ is the radius of the spherical colloid (lysozyme in our case). This results in a uniformly valid expression for the effective charge. The overshooting effect discussed in \cite{7} can then be understood at all $a \kappa$ for positively charged proteins or nanoparticles. The fully computed expression turns out to be subtle.

\section{Solution of the Poisson-Boltzmann equation for a 2--1 electrolyte}
 
The nanosphere bears a charge $Zq$ were $q$ is the elementary charge and $Z \!>\! 1$. The electrostatic potential $\varphi(r)$ between two spheres separated by a distance $r$ is scaled by $k_{\rm B} T$  where $k_{\rm B}$ is Boltzmann's constant and $T$ is the temperature: $\psi(r) \equiv q \, \varphi(r) / k_{\rm B} T$. If the concentration of 2--1 salt (MgBr$_2$ in the experiments to be discussed below) is $n$, the Debye screening length $\kappa ^{-1}$ is given by $\kappa^2 = 8 \pi Q I$ with ionic strength $I \!=\! 3 n$, the Bjerrum length $Q \!=\! q^{2} / D \, k_{\rm B} T$ where the permittivity $D$ is assumed to be uniform. The Poisson-Boltzmann equation then reads  
\begin{equation}
\Delta \Psi = \tfrac{1}{3} \, \kappa^2 \, \left( e^{\Psi} - e^{-2 \Psi} \right) \,,
\end{equation}
with boundary conditions 
\begin{eqnarray}
\frac{d \Psi}{dr} |_{r=a} &=& - \frac{Z \, Q}{a^2} \,, \\
\lim_{r \rightarrow \infty} \Psi(r)    &=& 0 \,.
\end{eqnarray}
The linearized version of eq (1) has the usual Debye-H\"{u}ckel solution
\begin{equation}
\Psi_0(r) = \frac{Z \, Q}{(1 + \kappa a)} \, \frac{e^{-\kappa(r-a)}}{r} \,.
\end{equation}
A pertubative solution to eq (1) is derived as follows (see Appendix 1 in ref 6; an error was made there -- the zero-order screening term was deleted -- but this is corrected here; fortunately, it turns out that errors incurred in the tables of ref 6 are within the margin of error). I seek a solution $\Psi = \Psi_0 + \Psi_1$ where $\Psi_1$ is uniformly smaller than $\Psi_0$ though $\Psi_0$ now has a higher effective charge $Z_ {\rm eff}$ instead of $Z$ owing to the divalent ion being substantially suppressed by the particle surface (see eq (1)). If we next scale the distance $r$ between the spheres by $\kappa$, $R \!=\! \kappa r$, we have
\begin{equation}
\frac{1}{R^2} \, \frac{d}{dR} \left( R^2 \, \frac{d\Psi_1}{dR} \right) = \Psi_1 + S(R) \,,
\end{equation}
with a source term 
\begin{equation}
S(R) \equiv - \tfrac{1}{2} \, \Psi_0^2 \,,
\end{equation}
and
\begin{eqnarray}
\Psi_0 &\equiv& \frac{B \, e^{-R}}{R} \,, \\
B      &\equiv& \frac{Z_{\rm eff} \, Q \, \kappa \, e^{\mu}}{(1 + \mu)} \,, \\
\mu    &\equiv& \kappa \, a \,.
\end{eqnarray}
Eq (5) is readily solved by quadrature. First, set $\Psi_1 \!\equiv\! f(R) / R$ which leads to
\begin{equation}
f^{\prime \prime} - f = R \, S(R) \,.
\end{equation}
Then, set $f(R) \!\equiv\! w(R) \, e^{-R}$ yielding
\begin{equation}
w^{\prime \prime} -2 \, w^{\prime} = R \, e^R \, S(R) \,.
\end{equation}
Noting that $w^{\prime} \rightarrow$ 0 as $R \rightarrow \infty$, we can integrate eq (11) to get
\begin{equation}
w^{\prime}(R) = \tfrac{1}{2} \, B^2 \, e^{2R} \, \int\limits_{R}^{\infty} \!\! dR^{\prime} \; \frac{e^{-3R^{\prime}}}{R^{\prime}} \,.
\end{equation}
Another integration gives
\begin{equation}
w(R) = - \tfrac{1}{2} \, B^2 \, \int\limits_{R}^{\infty} \!\! dR^{\prime\prime} \; e^{2R^{\prime\prime}} \,
\int\limits_{R^{\prime\prime}}^{\infty} \!\! dR^{\prime} \; \frac{e^{-3R^{\prime}}}{R^{\prime}} \,,
\end{equation}
since $w \rightarrow$ 0 as $R \rightarrow \infty$. I note that $w$ is negative as it should be. Eq (13) may be re-expressed in terms of the exponential integral
\begin{equation}
E_1(u) \equiv \int\limits_{u}^{\infty} \!\! dt \; \frac{e^{-t}}{t} \,,
\end{equation}
yielding
\begin{equation}
w(R) = - \tfrac{1}{4} \, B^2 \, \left[ E_1(R) - e^{2R} \, E_1(3R) \right] \,.
\end{equation}
We thus have that $\Psi_1(R) \!=\! w(R) \, \exp(-R) / R$ so eq (2) becomes
\begin{equation}
\left. \frac{d\Psi}{dR} \right|_{R=\mu} = - \frac{Z \, Q}{\kappa \, a^2} = \left. \frac{d\Psi_0}{dR} \right|_{R=\mu} + \left. \frac{d\Psi_1}{dR} \right|_{R=\mu} 
= - \frac{Z_{\rm eff} \, Q}{\kappa \, a^2} + \left. \frac{d\Psi_1}{dR} \right|_{R=\mu}
\end{equation}
The second term on the right-hand side of eq (16) is rewritten in terms of exponential integrals with the help of eqs (13) and (15)
\begin{equation}
\left. \frac{d\Psi_1}{dR} \right|_{R=\mu} = \tfrac{1}{4} \, B^2 \, \left[ \frac{e^{\mu}}{\mu} \, E_1(3\mu)
- \frac{e^{\mu}}{\mu^2} \, E_1(3\mu) + \frac{(\mu+1) \, e^{-\mu}}{\mu^2} \, E_1(\mu) \right] \,,
\end{equation}
which is always positive. Hence, we have $Z \!=\! Z_{\rm eff} - \lambda \, Z_{\rm eff}^2$ or $Z_{\rm eff} \!=\! Z + \lambda \, Z^2$ correct to ${\cal O}(Z^2)$ where
\begin{eqnarray}
\lambda &\equiv& \tfrac{1}{4} \, Q \, \kappa \, \left[ \frac{(\mu-1) \, e^{3 \mu}}{(\mu + 1)^2} \, E_1(3\mu)
+ \frac{e^{\mu}}{(\mu+1)} \, E_1(\mu) \right] \nonumber \\
&\equiv& \tfrac{1}{4} \, Q \, \kappa \, g(\mu) \,.
\end{eqnarray}
The asymptotic expansion of $E_1(\mu)$ at large $\mu$ is well known to be virtually useless \cite{9}. In effect, it is only as $\mu$ becomes exceedingly large ($\mu \!=\! {\cal O}(100)$) that eq (18) agrees with eq (3.16) computed by Tellez and Trizac \cite{7}. Obviously we need to use the full expression for $g(\mu)$ in practical calculations.
 
Developing a series expansion of the effective charge at low ionic strength ($a \kappa \!\ll\! 1$) is, however, straightforward. The leading term is of interest for it does not depend on $a$
\begin{equation}
Z_{\rm eff} = Z \, \left( 1 + \frac{Z \, Q \, \kappa \, \ln(3)}{4} \right) \,.
\end{equation}
This is a useful estimate for proteins and nanocolloids in the case $Z = {\cal O}(10)$ and $Q \kappa = {\cal O}(0.1)$ say. 

\section{Application to lysozyme in MgBr$_2$}

As in previous work \cite{6}, the radius of the lysozyme is set $a \!=$ 1.7 nm and the Bjerrum length $Q \!=$ 0.71 nm at room temperature ($T \!=$ 298 K). Kuehner {\em et al.} established the charge $Z \!=$ 7 at pH = 7.5 for lysozyme in a NaCl solution by titration \cite{10}. I assume this is also the bare charge for lysozyme in the case at hand. The theoretical second virial coefficient
\begin{equation}
B_2 = 2 \pi \, \int\limits_{0}^{\infty} \!\! dr \; r^2 \, \left( 1 - e^{-U(r) / k_{\rm B} T} \right) \,,
\end{equation}
is connected to the experimental second virial coeficient $B_{\rm exp}$ via $B_{\rm exp} = N_{\rm av} B_2 / M^2$ \cite{11} where $N_{\rm av}$ is Avogadro's number and $M$ is the molar mass (14.3 kg / mol for lysozyme). The hard sphere coefficient $B_{\rm HS} = 16 \pi a^{3} / 3 \!=$ 82 nm$^3$. Hence, if the lysozyme molecule were a sphere, the experimental hard-sphere value would be $B_{\rm exp, HS} \!=$ 2.41 $\times$ 10$^{-4}$ mol ml/g$^2$.

Tessier {\em et al.} measured $B_{\rm exp}$ of lysozyme in MgBr$_2$ solutions at pH = 7.8 by self-interaction chromatography \cite{4}. By contrast, Guo {\em et al.} had already determined $B_{\rm exp}$ by static light scattering in 1999 \cite{5}. The two sets are shown in Table 1. The second virial coefficient is probably difficult to measure accurately when the protein molecules attract each other, which may rationalize the disparity between the two methods. Both methods, however, establish that there is a minimum at about 0.3 M MgBr$_2$.

\begin{table}
\centering
\begin{tabular}{| c || c | c |}
MgBr$_2$ (M) & \multicolumn{2}{c |}{$B_{\rm exp}$ (10$^{-4}$ mol ml/g$^2$)} \\
             & SIC & SLS \\
\hline
\hspace*{3pt} 0.10 \hspace*{3pt} & \hspace*{8pt} -2.30  \hspace*{8pt} & \hspace*{3pt} -2.40 \hspace*{3pt} \\
\hspace*{3pt} 0.20 \hspace*{3pt} & \hspace*{8pt} -5.00  \hspace*{8pt} & \hspace*{3pt} -     \hspace*{3pt} \\
\hspace*{3pt} 0.30 \hspace*{3pt} & \hspace*{8pt} -6.14  \hspace*{8pt} & \hspace*{3pt} -4.50 \hspace*{3pt} \\
\hspace*{3pt} 0.43 \hspace*{3pt} & \hspace*{8pt} -5.24  \hspace*{8pt} & \hspace*{3pt} -4.40 \hspace*{3pt} \\
\hspace*{3pt} 0.53 \hspace*{3pt} & \hspace*{8pt} -4.25  \hspace*{8pt} & \hspace*{3pt} -3.70 \hspace*{3pt} \\
\hspace*{3pt} 0.70 \hspace*{3pt} & \hspace*{8pt} -2.70  \hspace*{8pt} & \hspace*{3pt} -3.20 \hspace*{3pt} \\
\hspace*{3pt} 1.00 \hspace*{3pt} & \hspace*{8pt}  0.00  \hspace*{8pt} & \hspace*{3pt} -     \hspace*{3pt} \\
\end{tabular}
\caption{Experimental second virial coefficient $B_{\rm exp}$ as a function of the magnesium bromide concentration.
Self-interaction chromatography (SIC) \cite{4}; Static light scattering (SLC) \cite{5}.}
\end{table}

Next, I shall use the comprehensive chromatography data to compute the number of magnesium ions using 
\begin{equation}
\frac{U(r)}{k_{\rm B} T} =  \left\{
\begin{array}{cl}
\infty                 & \hspace*{10pt} 0 \leq r \leq 2a \\
U_{\rm DH} - U_{\rm A} & \hspace*{5pt} 2a \leq r < 2a + \delta a \\
U_{\rm DH}             & \hspace*{11pt} r \geq 2a + \delta a \\
\end{array}
\right.
\end{equation}
which is analogous to the total interaction potential introduced in \cite{7} with Debye-H\"{u}ckel potential
\begin{equation}
U_{\rm DH}(r) = 2 \, a \, \xi \, \frac{e^{- \kappa (r-2a)}}{r} \,,
\end{equation}
where
\begin{equation}
\xi \equiv \frac{Q \, Z^2_{\rm eff}}{2 a \, (1 + \mu)^2} \,, 
\end{equation}
is the coupling parameter of the renormalized nonlinear Poisson-Boltzmann interaction. The depth of the adhesive well is $U_{\rm A}$ and it's thickness is $\delta a$. An accurate estimate of the second virial coefficient $B_2$ is computed in ref \cite{6}:
\begin{eqnarray}
\frac{B_2}{B_{\rm HS}} &=& 1 + \tfrac{3}{8} \, J \,, \\
J         &=&   J_1 - \left( e^{U_{\rm A}} - 1 \right) \, J_2 \,, \\
J_1    &\simeq& \frac{4 \, (\mu + \tfrac{1}{2}) \, \xi}{\mu^2} \, \left( 1 - \tfrac{1}{2} \, \alpha \, \xi \right) \,, \\
J_2    &\simeq& 2 \delta \left[ e^{-\xi} + (1 + \tfrac{\delta}{2})^2 \, \exp(- \frac{\xi}{1+\frac{\delta}{2}} \, e^{-\mu \, \delta}) \right] \,, \\
\alpha &\equiv& \left( e^{- \xi} - 1 + \xi \right) / \xi^2 \,.
\end{eqnarray}
In the case of lysozyme in a NaCl solution $\delta \!=$ 0.079, $U_{\rm A} \!=$ 3.70 and $\delta \, e^{U_{\rm A}} \!=$ 3.20 \cite{6}. It is assumed that these values pertain to lysozyme -- MgBr$_2$ solutions also. 
 
Table 2 is derived as follows. First, the effective charge $Z_{\rm eff}$ is computed from Poisson-Boltzmann electrostatics by numerically evaluating the function $g(\mu)$ (eq (18)). This pertains to the case when no magnesium ions are assumed to be bound to the lysozyme. The numerical calculation of the exponential integral is well known to be notoriously nontrivial but a powerful representation has been presented by applied mathematicians \cite{12}.

\begin{table}
\centering
\begin{tabular}{c || c | c | c | c | c | c | c}

MgBr$_2$ (M)               & \hspace*{1pt} 0.10 \hspace*{1pt}  & \hspace*{1pt}  0.20 \hspace*{1pt}  & \hspace*{1pt}  0.30 \hspace*{1pt} 
                           & \hspace*{1pt} 0.43 \hspace*{1pt}  & \hspace*{1pt}  0.53 \hspace*{1pt}  & \hspace*{1pt}  0.73 \hspace*{1pt} 
                           & \hspace*{1pt} 1.00 \hspace*{1pt}  \\

$\mu$                      & \hspace*{1pt} 3.06 \hspace*{1pt}  & \hspace*{1pt}  4.33 \hspace*{1pt}  & \hspace*{1pt}  5.31 \hspace*{1pt} 
                           & \hspace*{1pt} 6.35 \hspace*{1pt}  & \hspace*{1pt}  7.05 \hspace*{1pt}  & \hspace*{1pt}  8.10 \hspace*{1pt} 
                           & \hspace*{1pt} 9.69  \hspace*{1pt}  \\

$g(\mu)$                   & \hspace*{1pt} 0.0759  \hspace*{1pt} & \hspace*{1pt} 0.0466  \hspace*{1pt} & \hspace*{1pt} 0.0320  \hspace*{1pt}
                           & \hspace*{1pt} 0.0237  \hspace*{1pt} & \hspace*{1pt} 0.01985 \hspace*{1pt} & \hspace*{1pt} 0.01559 \hspace*{1pt}
                           & \hspace*{1pt} 0.01135 \hspace*{1pt} \\

$Z_{\rm eff}$              & \hspace*{1pt} 8.2  \hspace*{1pt}  & \hspace*{1pt}  8.0  \hspace*{1pt}  & \hspace*{1pt}  7.9  \hspace*{1pt} 
                           & \hspace*{1pt} 7.8  \hspace*{1pt}  & \hspace*{1pt}  7.7  \hspace*{1pt}  & \hspace*{1pt}  7.6  \hspace*{1pt} 
                           & \hspace*{1pt} 7.6  \hspace*{1pt}  \\

$\xi_{\rm eff}$            & \hspace*{1pt} 0.852 \hspace*{1pt}  & \hspace*{1pt}  0.471 \hspace*{1pt}  & \hspace*{1pt}  0.328 \hspace*{1pt} 
                           & \hspace*{1pt}  0.235 \hspace*{1pt}  & \hspace*{1pt}  0.191 \hspace*{1pt}  & \hspace*{1pt}  0.146 \hspace*{1pt} 
                           & \hspace*{1pt}  0.106 \hspace*{1pt}  \\

$B_2 / B_{\rm HS}$         & \hspace*{1pt}  -0.95 \hspace*{1pt}  & \hspace*{1pt}  -2.07 \hspace*{1pt}  & \hspace*{1pt}  -2.55 \hspace*{1pt} 
                           & \hspace*{1pt}  -2.17 \hspace*{1pt}  & \hspace*{1pt}  -1.76 \hspace*{1pt}  & \hspace*{1pt}  -1.12 \hspace*{1pt} 
                           & \hspace*{1pt}   0.0  \hspace*{1pt}  \\

$\xi_{\rm num}$            & \hspace*{1pt}  0.839 \hspace*{1pt}  & \hspace*{1pt}  0.490 \hspace*{1pt}  & \hspace*{1pt}  0.357 \hspace*{1pt} 
                           & \hspace*{1pt}  0.508 \hspace*{1pt}  & \hspace*{1pt}  0.691 \hspace*{1pt}  & \hspace*{1pt}  1.945 \hspace*{1pt} 
                           & \hspace*{1pt}  2.15  \hspace*{1pt}  \\

$Z_{\rm eff, num}$         & \hspace*{1pt}   8.1 \hspace*{1pt}  & \hspace*{1pt}   8.2 \hspace*{1pt}  & \hspace*{1pt}   8.2 \hspace*{1pt} 
                           & \hspace*{1pt}  11.5 \hspace*{1pt}  & \hspace*{1pt}  14.6 \hspace*{1pt}  & \hspace*{1pt}  27.7 \hspace*{1pt} 
                           & \hspace*{1pt}  34.2 \hspace*{1pt}  \\

bound Mg$^{2+}$            & \hspace*{1pt}   0  \hspace*{1pt}  & \hspace*{1pt}   0  \hspace*{1pt}  & \hspace*{1pt}   0  \hspace*{1pt} 
                           & \hspace*{1pt}   2  \hspace*{1pt}  & \hspace*{1pt}  3-4 \hspace*{1pt}  & \hspace*{1pt}  10  \hspace*{1pt} 
                           & \hspace*{1pt}  13  \hspace*{1pt}  \\

\end{tabular}
\caption{Number of magnesium ions bound to lysozyme computed as outlined in the text. The function $g(\mu)$ is calculated numerically with the help of the procedure in ref.\cite{12} and $Z_{\rm eff} \!=\! Z + \lambda \, Z^2$. The purely electrostatic coupling constant $\xi_{\rm eff}$ is given by eq (23). When the attractive potential between two lysozyme spheres is switched on, $\xi_{\rm num}$ is computed from eqs (24)-(28) and $Z_{\rm eff, num}$ from eq (23).}
\end{table}

Next, the coupling parameter $\xi$ is supposed adjustable in view of Mg$^{2+}$ binding to the surface and is computed numerically by imposing the experimental values of $B_2 / B_{\rm HS}$ \cite{4} in eq (24). Eq (23) then yields the corresponding numerically adjusted $Z_{\rm eff, num}$. The number of bound Mg ions is simply ($Z_{\rm eff, num} - Z_{\rm eff}) / 2$. The number of bound ions in the densitometry experiments at 1 M MgCl$_2$ was 4 at pH = 3.0 and 6 at pH = 4.5, which would lead to a tentative estimate of 10 at pH = 7.5. At 0.5 M MgCl$_2$, this number was 3 at pH = 3.0.

The pH = 7.8 of the solutions used by Tessier {\em et al.} \cite{4} is not identical with the pH = 7.5 of the solutions used by Kuehner {\em et al.} \cite{10}. In eqs (25) and (27) the parameters $\delta$ and $U_{\rm A}$ pertain to the latter pH \cite{6}. However, the small disparity in pH does not affect these conclusions. From ref. \cite{10} one may infer  $\Delta Z / \Delta {\rm pH} \simeq$ -0.93  so that $\Delta Z \simeq$ -0.28 at pH = 7.8. This leads to $\Delta Z_{\rm eff} \simeq$ 0.3 in the range of 0.1-0.3 M MgBr$_2$. A relative change of the effective charge by 10\% does not alter the number of bound magnesium ions in the interval (= zero).

It is also of interest to consider the properties of the layers of Mg$^{2+}$ ions at the two highest MgBr$_2$ concentrations of Table 2. The surface area of lysozyme is 36.3 nm$^2$. If the ions were spread evenly across the protein surface, the typical distance between neighboring ions would be $d \!=$ 2 nm. The relevant electrostatic energy $4 Q \, k_{\rm B} T / d = \cal{O}$(1) is not weak let alone if we were to include interactions beyond nearest neighbors. Hence, the nonconvalently bound ions would appear to be spread uniformly across the lysozyme surface.

\section{Discussion}

We now have a mean-field theory of lysozyme in which its properties in monovalent salt (second viral coefficient \cite{6}, concentration dependence of the osmotic pressure \cite{6}, crystallization \cite{13} and diffusivity \cite{14}) and in divalent salt (this work) are well explained by Poisson-Boltzmann electrostatics together with a well-defined adhesive well ($U_{\rm A}$, $\delta a$ with $U_{\rm A}$ and $\delta$ fixed parameters) which is independent of electrostatics (ionic strength). Moreover, Prinsen and I have argued on general grounds that traditional dispersion forces cannot even begin to describe the second viral as a function of the ionic strength \cite{6}. Very extensive computational work \cite{15} on lysozyme in MgBr$_2$ solutions is in stark disagreement with the experimental data \cite{4,5}: the predicted second virial coefficient decreases rapidly with the concentration of magnesium bromide. Accordingly, without a posited adhesive well, it is difficult to see how the minimum in the data for $B_2$ can be explained at all. Note that the computational physics of biopolymers is still in a state of considerable flux \cite{16}.

\section{Concluding remarks}

It is expected that no magnesium ions are bound to the lysozyme at low concentrations and this is well borne out by the first three entries in Table 2. Beyond the minimum in $B_2$, the second virial coefficient measured by Tessier {\em et al.} imposes a value of 13 bound magnesium ions at 1 M MgBr$_2$ compared with a tentative extrapolation of 10 bound magnesium ions by densitometry \cite{3}. Accordingly, it would be of interest to perform new measurements at the appropriate pH in a full range of ionic strengths to see how well the current theory applies.

\end{document}